\documentclass[10pt]{iopart}
\eqnobysec
\def\Schrodinger{{Schr\"odinger} }

\def\ie{{\it i.e., }}
\def\half{{1\over 2}}
\begin{document}
\title{\Schrodinger representation for the polarized Gowdy model}
\medskip
\author
{C.~G.~Torre}
\smallskip
\address{
Department of Physics,
Utah State University,
Logan, UT 84322-4415
USA}

\begin{abstract}
The polarized ${\bf T}^3$ Gowdy model is, in a standard gauge, characterized by a point particle degree of freedom and a scalar field degree of freedom obeying a linear field equation on ${\bf R}\times{\bf S}^1$.  The Fock representation of the scalar field has been well-studied. Here we construct the \Schrodinger  representation for the scalar field at a fixed value of the Gowdy time in terms of square-integrable functions on a space of distributional fields with a Gaussian probability measure. We show that ``typical'' field configurations are slightly more singular than square-integrable functions on the circle. For each time the corresponding \Schrodinger representation is unitarily equivalent to the Fock representation, and hence all the \Schrodinger representations are equivalent.  However, the failure of unitary implementability of time evolution in this model 
manifests itself in the mutual singularity of the Gaussian measures at different times.
\end{abstract}

\section{Introduction}

The Gowdy class of vacuum spacetimes \cite{Gowdy1974} has for many years been an attractive toy model for investigations in quantum gravity; see {\it e.g.}, references [2--9].
%\cite{Misner1973, Berger, Husain1987, Husain1989, Pierri, Corichi2002, CGT2002}.
 These spacetimes arise from a symmetry reduction of the Einstein theory by an Abelian 2-parameter group and define a large class of inhomogeneous cosmological models featuring a big bang. From the point of view of quantum gravity the ``polarized'' sub-class of models define exactly soluble quantum field theories  owing to the infinite number of degrees of freedom remaining after the symmetry  reduction. Thus these models can be used to explore intrinsically field-theoretic effects which may feature in a complete quantum theory of gravity. 

The traditional quantization of the polarized ${\bf T}^3$ Gowdy model uses the fact that in an appropriate gauge the phase space of the theory is characterized by a point particle degree of freedom and a scalar field degree of freedom, the latter satisfying a linear field equation. The point particle degree of freedom is treated in the usual quantum mechanical way, while the scalar field is quantized via a Fock representation of the canonical commutation relations deriving from a natural notion of positive frequency solutions to the field equation. Here we investigate the corresponding \Schrodinger representation of the scalar field, \ie we construct a representation in terms of functions of the value of the scalar field at a fixed time.  There are several reasons why this representation is of interest.  Perhaps most importantly, the \Schrodinger representation is the setting for the original (and still relatively poorly understood) approach to canonical quantum gravity based upon functions of 3-metrics \cite{DeWitt1967, Kuchar1992}. It is also akin to the representation currently proposed for the full theory using the better understood ``loop quantization'', based upon functions of connections \cite{AA2004}. In loop quantum gravity the dynamical evolution of the theory is still not well-understood, while in the polarized Gowdy model the dynamical evolution is quite tractable and, as we shall see, has an important interplay with the \Schrodinger representation. Moreover, the polarized Gowdy model sits as an exactly soluble model within the larger class of unpolarized Gowdy models. The latter are self-interacting field theories and could be studied using perturbative methods starting from the polarized model. Functional integral approaches based upon the \Schrodinger representation have been one of the principal methods of rigorously constructing interacting quantum field theories \cite{GJ1987}. Finally, the \Schrodinger representation gives immediate physical insight into fundamental properties of the quantum field. In particular, the support of the measure used to define the representation reveals the mathematical behavior of ``typical'' field configurations, {\it i.e.}, typical quantum metrics.  This behavior is far more intricate for field theories than for quantum mechanical systems. 

The \Schrodinger representation for the polarized ${\bf T}^3$ Gowdy model is defined in terms of functions on a space of distributional field configurations representing the scalar field at a fixed value of the Gowdy time coordinate. This space of field configurations is equipped with a Gaussian probability measure determined by requiring equivalence with the usual Fock representation. We will show that the support of the measure is within a Sobolev space which is slightly more ``singular'' than that of square-integrable field values. This representation can be constructed for any value of the Gowdy time and the resulting representations are all unitarily equivalent. One of the novel, intrinsically field-theoretic features of this model is the failure of unitary implementability of time evolution \cite{Corichi2002, CGT2002}. This causes no difficulties in the mathematical construction and/or physical interpretation of the \Schrodinger representation. However, this feature does prevent the mutual continuity of the Gaussian measures at different times. Indeed, we show that the measures at different times are mutually singular.

\section{The classical Gowdy model}

The Gowdy model we will study arises by assuming spacetime is not flat, that its manifold is ${\cal M}={\bf R}^+\times {\bf  T}^3$ with spacelike ${\bf  T}^3$, and that there is an Abelian 2-parameter isometry group, $G={\bf T}^2$, with spacelike  orbits ${\bf T}^2\subset {\bf T}^3$ generated by a pair of commuting, hypersurface-orthogonal Killing vector fields.  We will use coordinates $(t,x,y,z)$ on $\cal M$, where $t>0$, and $(x,y,z)\in (0,2\pi)$. These coordinates are chosen such that the $t=const.$ surfaces foliate $\cal M$ with spacelike ${\bf T}^3$ hypersurfaces (coordinates $(x,y,z)$), and such that the Killing vector fields are ${\partial\over\partial y}$ and ${\partial\over\partial z}$.  In such coordinates
the metric $g$ on $\cal M$ can be put into the form \cite{RT1996}
\begin{equation}
\eqalign{
g=[&-(N^{\perp})^2 + e^{\gamma-\psi}(N^x)^2]dt^2
+ 2e^{\gamma-\psi}N^xdtdx
+ e^{\gamma-\psi}dx^2\cr
&+ \tau^2e^{-\psi}dy^2 + e^\psi dz^2.}
\end{equation}
The functions $N^\perp(t,x)$ and $N^x(t,x)$ determine the lapse function and shift vector for $G$-invariant spacelike foliations of $\cal M$. The function $\tau(t,x)>0$ defines the area of the orbits of $G$ and will be used as a time coordinate. The gradient of $\tau$ is assumed to be everywhere timelike.
  The functions $\gamma(t,x)$ and $\psi(t,x)$ are unrestricted.

A Hamiltonian formulation of the symmetry-reduced Einstein equations is given in \cite{RT1996}, and we shall use this formulation here.  Using a prime and dot to denote differentiation with respect to $x$ and $t$ respectively, the canonical action functional for the theory defined on ${\cal M}/G\approx {\bf R}^+\times {\bf S}^1$ is given by
\begin{equation}
\eqalign{
S&[N^\perp,N^x,\gamma,\tau,\psi,P_\gamma,P_\tau,P_\psi] =\int_{t_1}^{t_2}dt\,\int_0^{2\pi}
dx\,\Bigg(P_\gamma\dot\gamma +P_\tau\dot \tau + P_\psi\dot\psi 
\cr
 &-N^\perp e^{(\psi-\gamma)/2}\Big[-
P_\gamma P_\tau+2\tau^{\prime\prime}-
\tau^\prime\gamma^\prime
+{1\over2}\left(\tau^{-1}P_{\psi}^2+\tau\psi^{\prime2}\right)
\Big]\cr
&- N^x\Big[
-2P_\gamma^\prime 
+P_\gamma\gamma^\prime+P_\tau\tau^\prime+P_\psi\psi^\prime\Big]\Bigg),}
\end{equation}
which displays the canonical coordinates $(\gamma, \tau, \psi, P_\gamma, P_\tau, P_\psi)$ for the phase space $\Gamma$. The symplectic 2-form on $\Gamma$ is
$$
\Omega = \int_0^{2\pi} dx\, \left(\delta P_\tau\wedge \delta \tau + \delta P_\gamma\wedge \delta \gamma + \delta P_\psi \wedge \delta \psi\right).
$$
Variation of the action with respect to the lapse and shift functions $(N^\perp, N^x)$ yields the Hamiltonian and momentum constraints
\begin{equation}
H_\perp=0=H_x,
\end{equation}
where
the Hamiltonian and momentum constraint functions are given by:
\begin{equation}
H_\perp = e^{(\psi-\gamma)/2}\Big[-
P_\gamma P_\tau+2\tau^{\prime\prime}-
\tau^\prime\gamma^\prime
+{1\over2}\left(\tau^{-1}P_{\psi}^2+\tau\psi^{\prime2}\right)\Big],
\label{Hperp}
\end{equation}
\begin{equation}
H_x=-2P_\gamma^\prime
+P_\gamma\gamma^\prime+P_\tau\tau^\prime+P_\psi\psi^\prime.
\label{Hx}
\end{equation}
The restriction that the spacetime gradient of $\tau$ is timelike leads to a corresponding restriction on the phase space $\Gamma$:
\begin{equation}
P_\gamma <-|\tau^\prime|.
\end{equation}

We  will work in the gauge defined by
\begin{equation}
\tau/{\cal P}=t,\quad  P_\gamma = -{\cal P},\label{gauge1}
\end{equation}
where\footnote{We note that $\cal P$ has vanishing Poisson brackets with the 
constraint functions.}
\begin{equation}
{\cal P} = -{1\over 2\pi}\int_0^{2\pi} dx  P_\gamma>0.
\end{equation}
Preserving the gauge fixing conditions in time gives (modulo the constraints)
\begin{equation}
N^\perp =  e^{(\gamma-\psi)/2},\quad (N^x)^\prime = 0.
\end{equation}
The constraints become
\begin{equation}
H_\perp = e^{(\psi-\gamma)/2}\left[{\cal P}  P_\tau +\half({1\over t{\cal P}} P_\psi^2 + t{\cal P}\psi^{\prime2})\right]\approx 0,
\end{equation}
\begin{equation}
 -{\cal P}\gamma^\prime +  P_\psi\psi^\prime \approx 0,
\end{equation}
with solution
\begin{equation}
 P_\tau =  -\half{1\over{\cal P}}({1\over t{\cal P}} P_\psi^2 + t{\cal P} \psi^{\prime2}),
 \label{ptau}
\end{equation}
\begin{equation}
\gamma(x) = \gamma_0 + \int_0^x dx^\prime {1\over {\cal P}} P_\psi(x^\prime)\psi^\prime(x^\prime),
\label{gamma}
\end{equation}
and the remaining constraint
\begin{equation}
K=\int_0^{2\pi}dx\,   P_\psi\psi^\prime \approx 0.
\end{equation}
This is the only first-class constraint remaining after gauge fixing. $K$ generates a 1-parameter group of canonical transformations corresponding to a rigid translation in $x$, which is the coordinate freedom remaining after imposing (\ref{gauge1}). 

The conditions (\ref{gauge1}), (\ref{ptau}) and (\ref{gamma}) define a subspace $i\colon \tilde\Gamma\to \Gamma$; the induced symplectic structure is
\begin{equation}
i^*\Omega = -d h \wedge dt + d {\cal P}\wedge d {\cal Q} + \int_0^{2\pi}dx \, \delta P_\psi \wedge \delta \psi
\end{equation}
where 
\begin{equation}
{\cal Q} = -2\pi\gamma_0 - {t\over{\cal P}}h - \int_0^{2\pi} dx\, 
\int_0^x dx^\prime {1\over {\cal P}} P_\psi(x^\prime)\psi^\prime(x^\prime).
\end{equation}
and 
\begin{equation}
h = \int_0^{2\pi} dx\, \half({1\over t{\cal P}} P_\psi^2 + t{\cal P} \psi^{\prime2}).
\end{equation}
The quantity $h$ is therefore the Hamiltonian for the dynamics of the gauge-reduced system on $\tilde \Gamma$ with canonical chart $({\cal Q},{\cal P}; \psi,P_\psi)$, still subject to the constraint 
\begin{equation}
\int_0^{2\pi}dx\,   P_\psi\psi^\prime \approx 0.
\end{equation} 

We now make a convenient change of variables on $\tilde\Gamma$.  We define
\begin{equation}
\phi := \sqrt{\cal P}\psi,\quad P_\phi := {1\over \sqrt{\cal P}}P_\psi.
\end{equation}
and
\begin{equation}
Q = {\cal P Q} + \half\int_0^{2\pi} dx\,  P_\psi \psi,\quad P=\ln {\cal P} .
\end{equation}
Because
\begin{equation}
d{\cal P}\wedge d{\cal Q} + \int_0^{2\pi} dx\, \delta P_\psi\wedge \delta \psi
=dP\wedge dQ +  \int_0^{2\pi} dx\, \delta P_\phi\wedge \delta \phi
\label{ss}
\end{equation}
it is clear this is a  canonical transformation.

We are thus led to the following description of the dynamics after deparametrization. The phase space $\tilde\Gamma$ includes a point particle degree of freedom described by canonical variables $(Q,P)$ and a field degree of freedom described by canonical variables $(\phi,P_\phi)$. These variables are evolved in the time $t$ by the Hamiltonian(s)
\begin{equation}
H = \int_0^{2\pi} dx\, \half({1\over t }P_\phi^2 + t \phi^{\prime2}).
\end{equation}
The field degrees of freedom are subject to a first-class constraint
\begin{equation}
K=\int_0^{2\pi}dx\,  P_\phi\phi^\prime \approx 0.
\end{equation}
The Hamilton equations of motion (in the gauge $N^x=0$) are
\begin{equation}
\dot Q =   \dot  P = 0,
\end{equation}
and
\begin{equation}
\dot \phi = {1\over t} P_\phi,\quad \dot P_{\phi} =   t \phi^{\prime\prime},
\end{equation}
which imply
\begin{equation}
\ddot \phi + {1\over t}\dot \phi - \phi^{\prime\prime} = 0.
\label{phieq}
\end{equation}
The solutions of the equation of motion are 
\begin{equation}
Q(t) = Q_0,
\quad
{ P}(t) = {P}_0,
\end{equation}
\begin{eqnarray}
\phi(t)= &q + p \ln t\\
&+{1\over2\sqrt{2}} \sum_{n\neq 0}\left( a_n  H_0(|n|t)e^{-in x} + a^*_n  H^*_0(|n|t)e^{in x}\right),
\label{psi_soln}
\end{eqnarray}
\begin{equation}
P_\phi(t) =  t \dot \phi(t),
\end{equation}
Here $H_0$ is the zeroth order Hankel function of the second kind.  The symplectic structure (\ref{ss}) implies the following non-vanishing Poisson brackets:
\begin{equation}
[q,p] =1=[Q,P],\quad [a_n,a_m^*] = -i \delta_{nm}.
\end{equation}
The constraint $K=0$ is equivalent to
\begin{equation}
\sum_{n\neq0} n |a_n|^2 = 0.
\label{k0}
\end{equation}

\section{Quantization: the Fock representation}

Here we briefly review the by-now standard Fock space quantization of the model.  Define a Hilbert space ${\cal H}$ by
\begin{equation}
{\cal H} = L^2({\bf R}^2) \otimes {\cal F},
\end{equation}
where ${\cal F}$ is the symmetric Fock space built from the Hilbert space of square-summable complex sequences,
$
 \zeta_n, n=\pm1, \pm2, \ldots, 
$
\begin{equation}
\sum_{n\neq 0} |\zeta_n|^2 < \infty.
\end{equation}
Any $\Psi\in\cal F$ can be represented as an infinite sequence of complex sequences\footnote{Here $\psi_0$ is the ``vacuum amplitude'', $\psi_{m_1}$ is the amplitude for ``1-particle with momentum $m_1$'', {\rm etc.}  For convenience, we represent the entire sequence $\{\psi_k,\ k = \pm 1, \pm2, \ldots\}$ simply by the symbol $\psi_k$.}
\begin{equation}
\Psi = (\psi_0,\psi_{m_1},\psi_{m_1m_2},\ldots,\psi_{m_1\cdots m_k},\ldots),
\label{Psi}
\end{equation}
where $\psi_0\in {\bf C}$,
\begin{equation}
\psi_{m_1\cdots m_k} = \psi_{(m_1\cdots m_k)},
\label{symm}
\end{equation}
and
\begin{equation}
|\psi_0|^2 + \sum_{k=1}^\infty \, \sum_{m_1\cdots m_k\neq 0}|\psi_{m_1\cdots m_k}|^2 < \infty.
\label{norm}
\end{equation}

We now define operators $(\hat Q, \hat P, \hat \phi, \hat P_\phi)$ corresponding to the canonical variables we constructed above. 
The canonical pairs $(\hat Q,\hat P)$ and $(\hat q,\hat p)$  are represented as identity operators on $\cal F$ and are represented on (a dense domain in) $L^2({\bf R}^2)$ exactly as one would canonical coordinates and momenta for a particle moving in two dimensions, {\it e.g.},
\begin{equation}
\psi=\psi(x,y)\in L^2({\bf R}^2),
\end{equation}
\begin{equation}
\hat q\psi = x\psi,\quad \hat p\psi = {1\over i}\partial_x\psi,\quad \hat Q\psi=y\psi,\quad
\hat P\psi = {1\over i}\partial_y\psi.
\label{qprep}
\end{equation}
(Other equivalent representations are, of course, possible.) 
 The remaining degrees of freedom in the field $\hat \phi$ are represented as identity operators on $L^2({\bf R}^2)$ and are represented on the Fock space $\cal F$ as follows.  
Using the representation (\ref{Psi}) for $\Psi\in {\cal F}$, we define annihilation and creation operators for each $l\neq 0$ by
\begin{eqnarray}
\hat a_l\Psi &= (\psi_l, \sqrt{2} \psi_{lm_1},\sqrt{3}\psi_{lm_1m_2},\ldots)\\
\hat a_l^*\Psi &= 
(0, \psi_0\delta_{m_1l},\sqrt{2}\delta_{l(m_1}\psi_{m_2)},\sqrt{3}\delta_{l(m_1}\psi_{m_2 m_3)},\ldots). 
\end{eqnarray}
These operators (on their common domain) satisfy
\begin{equation}
\hat a_n^* = (\hat a_n)^\dagger,\quad [\hat a_n\hat ,a_m^*] = \delta_{nm} \hat I,
\label{a_alg}
\end{equation}
where $\hat I$ is the identity operator on $\cal H$. 
The quantum field $\hat \phi$ is defined as an operator-valued distribution on ${\bf R}^+\times {\bf S}^1$ using the operator representation of $(\hat q,\hat p,\hat a_n,\hat a_n^*)$ in the expansion (\ref{psi_soln}). 

This quantization just described satisfies the prescription 
\begin{equation}
\{{\rm Poisson\ Bracket}\}\leftrightarrow {1\over i}\ [{\rm commutator}]
\end{equation}
for the canonical coordinates and momenta.  As shown {\it e.g.}, in \cite{Wald1994}, quantizations of $\phi$ of this type can be placed in correspondence with a choice of scalar product (equivalently, a choice of complex structure) on the space of solutions to (\ref{phieq}).  This is tantamount to making a choice of ``positive frequency solutions'' to the field equation.  The choice of positive frequency being made here can be understood as follows.  As shown in \cite{RT1996} one can view the scalar field and its field equation as arising from a symmetry reduction of a massless scalar field restricted to a submanifold of $2+1$ dimensional Minkowski space.   It is straightforward to show that the notion of positive frequency ({\it i.e.}, choice of scalar product, complex structure) being used to define the Fock representation above is the same as the usual notion of positive frequency in this $2+1$ dimensional Minkowski space.

It has been shown that, within this Fock representation, it is not possible to represent the Heisenberg picture time evolution of the field operators using a unitary transformation. In other words, there is no operator $\hat U$ on the Fock space such that
\begin{equation}
\hat \phi(t_2) = \hat U^{-1} \hat \phi(t_1) \hat U.
\end{equation}
 Roughly speaking, time evolution of the field variables can be viewed as a deformation of the scalar product (complex structure, notion of positive frequency) used to define the Fock representation. This deformation is too ``large'' to be represented as a unitary transformation.  While this is a technical inconvenience ({\it e.g.}, the \Schrodinger {\it picture} of time evolution is not available), it is not clear what --- if any --- problem this really causes for the physical viability of the quantization.  In the context of the present investigation, we shall see that while this field-theoretic feature of the model has implications for the support of the measure in the \Schrodinger representations associated with different times, it does not cause any difficulties with the construction/interpretation of the \Schrodinger representation.

\section{The \Schrodinger representation}

The Fock representation for the spatially inhomogeneous modes of $\hat \phi$ is analogous to the number operator representation of the harmonic oscillator. The representation we build now is, using the same analogy, the position  wave function representation. The ingredients in the construction are (i) a space ${\cal S}^\prime$ of distributional field configurations, (ii) a Gaussian integration measure $d\mu$ on ${\cal S}^\prime$, (iii) a representation of the (spatially inhomogeneous modes of the) field and its conjugate momentum at a fixed time in terms of differential operators on $L^2({\cal S}^\prime, d\mu)$. We are adopting the approach of Glimm and Jaffe \cite{GJ1987}.

We want a representation in which the  field operator at a fixed time $t=T$, denoted by $\hat \phi(T)$, acts by multiplication. The spatially homogeneous mode of $\hat\phi(T)$ can be treated in the usual way for a point particle. We focus on the spatially inhomogeneous modes in what follows. Let ${\cal S}$ denote the pre-Hilbert space of smooth real functions $f$ on the circle with vanishing integral:
\begin{equation}
\int_{S^1} f = 0.
\end{equation}
We introduce the scalar product for $f,g\in {\cal S}$:
\begin{equation}
(f,g) = \int_{S^1} fg.
\end{equation}
We can write, for $f\in {\cal S}$,
\begin{equation}
f = \sum_{n\neq 0} f_n e_n,
\end{equation}
where
\begin{equation} 
e_n(x) = {1\over\sqrt{2\pi}} e^{-inx}\in {\cal S},\quad f_{-n} = f_n^*,\quad n\neq0
\end{equation}
and $\{f_n\}$ is a rapidly decreasing sequence of complex numbers, that is, 
\begin{equation}
\lim_{n\to\infty}n^p f_n = 0\quad \forall p>0.
\end{equation}
Thus $\cal S$ can be identified with the set of rapidly decreasing complex numbers and is a nuclear space, as described, {\it e.g.,} in \cite{RS, GJ1987}. 

Let ${\cal S}^\prime$ denote the real vector space of linear functions on the nuclear space $\cal S$ which are continuous in the Frechet topology.   This will be the ``quantum configuration space'' used to define the \Schrodinger representation. (We shall see that this configuration space can be reduced in size, however.)
Each $Q\in {\cal S}^\prime$ is determined by its Fourier components:
\begin{equation}
Q(f) = \sum_{n\neq 0} Q_n f_n,
\end{equation}
where
\begin{equation}
Q_n =  Q(e_n),\quad Q_{-n} = Q_n^*.
\end{equation}
The sequence $\{Q_n\}$ is slowly increasing, that is, the sequence of real numbers $\{|Q_n|\}$, $n=1,2,\dots$, increases no faster than a polynomial in $n$.

Given $\cal S$ and ${\cal S^\prime}$ we are in a position to construct an integration measure on  ${\cal S^\prime}$. As shown in \cite{GJ1987}, a Gaussian integration measure on ${\cal S}^\prime$ is uniquely determined by a choice of a continuous, positive definite bilinear form ${\bf C}\colon {\cal S}\times{\cal S} \to {\bf R}$,
\begin{equation}
f,g\in {\cal S}\to {\bf C}(f,g) = (f,Cg).
\end{equation}
Our strategy is to find the form of $C$ dictated by the quantization of \S 3 and define the representation space for the spatially inhomogeneous modes of the scalar field operator  as $L^2({\cal S^\prime}, d\mu)$, where $d\mu$ is the Gaussian measure defined by $C$.
A general formalism for finding the form of $C$ appropriate to a given Fock representation can be found in \cite{Corichi2004}. A straightforward application of such techniques to the case at hand yields, for $f\in{\cal S}$,
\begin{equation}
Cf = {1\over8}\sqrt{\pi\over 2}\sum_{n\neq 0} |H_0(|n|T)|^2 f_ne^{-inx}. 
\label{Cdefn}
\end{equation}
Because $|H_0(z)|$ decays like ${1\over \sqrt{z}}$ as $z\to\infty$ the Fourier components of $Cf$ are rapidly decreasing so that  $C\colon {\cal S}\to {\cal S}$. Because $|H_0(|n|T)|$ is bounded and positive definite for fixed $T$ and $n>0$ it follows that ${\bf C}$ is continuous and  positive definite on $\cal S\times {\cal S}$ as it should be.  We remark that both the covariance and its associated measure depend upon the choice of time $T$. We denote the Gaussian integration measure defined by the covariance (\ref{Cdefn}) by $d\mu_{\scriptscriptstyle T}$.

What is the nature of the Gaussian integration defined by the covariance (\ref{Cdefn})? On  the cylinder set  defined by $\{Q_{-N},Q_{-N+1},\dots,Q_{-1},Q_1,\dots,Q_{N-1}, Q_N\}\in {\cal R}$ the covariance defines a measure  given in terms of the Lebesgue measure $dQ$  by
\begin{equation}
d\mu_{\scriptscriptstyle T}\Big|_{\cal R} = \prod_{j=1}^N {4\over \pi^2 |H_0(|j|T)|^2} \exp\left\{{-{4\over\pi|H_0(|j|T)|^2}|Q_j|^2}\right\} dQ_j dQ_j^*.
\label{Rmeasure}
\end{equation}
%Formally, the Gaussian measure $d\mu$ on ${\cal S}^\prime$ can be expressed as
%$$
%d\mu =\prod_{n\neq0}{2\over \pi^2 |H_0(|n|T)|^2}\exp\left\{-{2\over\pi|H_0(|n|T)|^2}|Q_n|^2\right\}.
%$$

The \Schrodinger representation 
is defined upon the Hilbert space $ L^2({\cal S}^\prime,d\mu_{\scriptscriptstyle T})$. We still have to represent the field operators.  It is convenient to work with the Fourier components of the field operators at time $T$. Using the pairing $\langle\cdot,\cdot \rangle$ between ${\cal S}^\prime$ and $\cal S$, we have
\begin{equation}
\hat \phi_n = \langle\hat  \phi(T),e_{-n}\rangle,\quad \hat P_n =  \langle\hat P_\phi(T),e_{n}\rangle.
\end{equation}
Let us consider the following representation:
\begin{eqnarray}
\hat \phi_n \Psi(Q) &= Q_n\Psi(Q_n),\\ 
\hat P_n\Psi(Q) &= \left({1\over i}{\partial \Psi(Q)\over\partial Q_n} -|n|T{H_1^*(|n|T)\over H_0^*(|n|T)} Q_{-n}
\Psi(Q)\right)
\end{eqnarray} 
We need to check the commutation relations 
\begin{equation}
[\hat \phi_n,\hat P_m]=i\delta_{nm} \hat I,\quad [\hat \phi_n,\hat \phi_m]=0=[\hat P_n,\hat P_m],
\end{equation}
and the reality conditions
\begin{equation}
\hat \phi_n^\dagger =\hat  \phi_{-n},\quad \hat P_n^\dagger = \hat P_{-n}.
\end{equation}
The commutation relations easily follow from the fact that $\hat P$ differs from the operation of differentiation with respect to $Q$ by the gradient of a function of $Q$. Now consider the reality conditions. The conditions for $\hat \phi_n$ are obviously satisfied. For $\hat P_n$ we have
\begin{equation}
\eqalign{
\int d\mu_{\scriptscriptstyle T} \Phi^* \hat P_n\Psi &= \int d\mu_{\scriptscriptstyle T}\Big[({1\over i}\partial_{-n}\Phi)^*\Psi
+ {4\over \pi i |H_0(|n|T)|^2}Q_{-n}\Phi^*\Psi\cr 
&\phantom{= \int d\mu_{\scriptscriptstyle T}\Big[({1\over i}}- |n|T {H_1^*(|n|T)\over H_0^*(|n|T)} Q_{-n}\Phi^*\Psi\Big]\cr
&=\int d\mu_{\scriptscriptstyle T}\Big[({1\over i}\partial_{-n}\Phi)^*\Psi - |n|T{ H_1(|n|T)\over H_0(|n|T)} Q_{-n}\Phi^*\Psi\Big]\cr
&=\int d\mu_{\scriptscriptstyle T}\{({1\over i}\partial_{-n}- |n|T{ H^*_1(|n|T)\over H^*_0(|n|T)} Q_{n})\Phi\}^*\Psi\cr
&=\int d\mu_{\scriptscriptstyle T} (\hat P_{-n}\Phi)^*\Psi.
}
\end{equation}
Here we used the Wronskian identity
\begin{equation}
H_0(z) H_1^*(z) - H_0^*(z) H_1(z) = {4\over i\pi z}.
\end{equation}

This representation for the field operators is such that 
\begin{equation}
\hat a_n = {\sqrt{\pi}\over 2}H_0^*(|n|T) \partial_{-n},
\end{equation}
and
\begin{equation}
\hat a^*_n=\hat a^\dagger_n = -{\sqrt{\pi}\over 2}H_0(|n|T) \partial_n + {2\over\sqrt{\pi}}{1\over H_0^*(|n|T)} Q_{-n}.
\end{equation}
Note, in particular, the ``vacuum state'' is  $\Psi[Q]=1$ in this representation so the Gaussian measure defines the probability distribution of the field values (at time $T$) for this state. 

We remark that a parallel construction can be used to construct the \Schrodinger ``momentum representation'' in which the momentum operators at a given time are diagonal.  We shall not pursue this here.

Equivalence of the \Schrodinger representation with the Fock representation (see below) in conjunction with results of \cite{CGT2002} imply the field operators $\phi$ and $P_\phi$ can be defined as self-adjoint operators. One thus has the usual probability interpretation, {\it e.g.}, given a set ${\cal V}\subset {\cal  S}^\prime$ and a wave function $\Psi\in L^2({\cal S^\prime}, d\mu_{\scriptscriptstyle T})$, the probability ${\bf P}({\cal V})$ for finding $\phi(T)$ with ``values'' in $\cal V$ is given by
\begin{equation}
{\bf P}({\cal V}) = \int_{\cal V} d\mu_{\scriptscriptstyle T}[Q]\, |\Psi[Q]|^2.
\end{equation}

Physical states of the Gowdy model must satisfy a quantum form of the constraint (\ref{k0}).
As discussed, {\it e.g.}, in \cite{CGT2002}, this amounts to defining physical states as those belonging to  the kernel of 
\begin{equation}
\hat K=\sum_{n\neq0} n \hat a_n^\dagger a_n.
\end{equation}
 This 
space of states is  a Hilbert subspace spanned by the subset of the number operator basis with vanishing total field momentum. One can simply identify these states with their \Schrodinger representatives to obtain the space of physical states. It is straightforward to check that this is equivalent to defining physical states as functions $\Psi[Q]$ which are invariant under the group of phase transformations
\begin{equation}
Q_n\to e^{in\alpha} Q_n,\quad \alpha\in{\bf R}.
\end{equation}
Because the group is compact, the physical state wave functions are still square-integrable so this restriction will not cause any complications in what follows. 

We have seen that for each time $t=T$ there is a Gaussian integration measure $d\mu_{\scriptscriptstyle T}$ and a corresponding \Schrodinger representation on $L^2({\cal S^\prime}, d\mu_{\scriptscriptstyle T})$.  If we define a linear transformation which identifies the occupation number states in $\cal F$ with those in $L^2({\cal S^\prime}, d\mu_{\scriptscriptstyle T})$ it is straightforward to see this transformation defines a unitary equivalence of the Fock and \Schrodinger representations (see Theorem 6.3.4 in \cite{GJ1987}).   \Schrodinger representations constructed at any two times are therefore also equivalent.   This equivalence of the different \Schrodinger representations may seem surprising at first sight, given the failure of unitary implementability of time evolution. However, the latter result arises because time evolution for this model can be viewed as a non-trivial change in the complex structure which defines the Fock representation.  The \Schrodinger representations at different times all correspond to a fixed choice of complex structure, {\it i.e.,} a single Fock representation and are therefore equivalent.  Still, the failure of implementability does have implications for the \Schrodinger representation: it prevents the mutual continuity of the Gaussian measures constructed at different times \cite{Shale, Simon2}. Indeed, in the next section we shall show that  the one parameter family of Gaussian measures we have constructed are all mutually singular. 

\section{Support of the measure and regularity properties of the field}

The \Schrodinger representation is defined in terms of functions on the space of distributions ${\cal S}^\prime$, which suggests typical field configurations are rather singular. On the other hand, Gaussian measures are typically supported on a  smaller space than ${\cal S}^\prime$. For example, the Wiener measure for paths in one dimension is supported within a space of H\"older continuous paths.  Here we will give some results on the support of the probability measure $\mu_{\scriptscriptstyle T}$ featuring in the \Schrodinger representation. This will, in particular, give us some information about the regularity properties of typical field configurations.

We will apply the Minlos theorem in the form given in \cite{Yamasaki} (see also \cite{Simon}) to our Gaussian measure. This theorem gives the following information. Let $\bf  H$ be a Hilbert space with norm $||\cdot||$ and let $A$ be a Hilbert-Schmidt operator on $\bf H$. Define the norm $||\cdot||_A = ||A\cdot||$. If the covariance is continuous in $||\cdot||_A$, so that
\begin{equation}
(f,Cf) \leq {\rm const.} ||f||_A^2,
\end{equation}
 then the corresponding Gaussian measure lies in the topological dual ${\bf H}^\prime$.  

For our Hilbert spaces we will use a class of Sobolev spaces. Recall that for any $s\in {\bf R}$ the Sobolev spaces $H^s({\bf S}^1)$ can be defined in terms of the Fourier transform of $Q\in{\cal S}^\prime$ as \cite{Taylor1996}:
\begin{equation}
H^s({\bf S}^1)=\left\{Q\in {\cal S}^\prime\,|  \sum_n(1+n^2)^{s}|Q_n|^2 <\infty\right\}.
\end{equation}
The topological dual of $H^s$ is $H^{-s}$. 

In the Minlos theorem choose ${\bf H} = H^\epsilon({\bf S}^1)$, where $\epsilon>0$.  Define the linear operator $A$ on ${\bf H}$ by
\begin{equation}
(AQ)_n = {Q_n\over (1+n^2)^{{1\over 4} + {\epsilon\over 2}}}.
\end{equation}
It is easy to check that $A$ is Hilbert-Schmidt.  The norm defined by $A$ is the norm for $H^{-{1\over 2}}({\bf S}^1)$:
\begin{equation}
||Q||_A^2 = \sum_{n\neq 0} (1+n^2)^{-{1\over 2}} |Q_n|^2.
\end{equation}
We next check that the covariance (\ref{Cdefn}) is continuous in this norm.
We have that
\begin{equation}
(1+n^2)^{-{1\over 2}} \geq {{\rm const.}\over |n|},
\end{equation}
And we have \cite{Watson}
\begin{equation}
|H_0(z)|^2 \leq {{\rm const.}\over |z|},
\end{equation}
so that for any fixed $T>0$
\begin{equation}
\sum_{n\neq0} |H_0(|n|T)|^2 |f_n|^2 \leq {\rm const.} \sum_{n\neq0}(1+n^2)^{-{1\over 2}} |f_n|^2
\end{equation}
as desired.
We conclude that the Gaussian measure lies in $H^{-\epsilon}({\bf S}^1)$ for any $\epsilon>0$.  

We remark that the subspace $L^2({\bf S}^1)=H^0({\bf S}^1)\subset H^{-\epsilon}({\bf S}^1)$ has measure zero. This can be seen explicitly as follows.  Consider the function
\begin{equation}
\chi_\alpha[Q] =  \exp\left\{-{4\alpha\over \pi}\sum_{n=1}^\infty|Q_n|^2\right\},\quad \alpha \geq 0.
\end{equation}
If $Q\in L^2({\bf S}^1)$ then  $\lim_{\alpha\to0}\chi_\alpha=1$; otherwise $\lim_{\alpha\to0}\chi_\alpha=0$. Thus $\lim_{\alpha\to0}\chi_\alpha$ is the characteristic function on the set of square-integrable functions. We have 
\begin{eqnarray}
\mu_{\scriptscriptstyle T}(L^2({\bf S}^1))& =\lim_{\alpha\to0} \int d\mu_{\scriptscriptstyle T} \chi_\alpha[Q]\\ 
&=\lim_{\alpha\to0} \lim_{N\to\infty} \int d\mu_{\scriptscriptstyle T}  \exp\left\{-{4\alpha\over \pi}\sum_{n=1}^N|Q_n|^2\right\}.\label{IN}
\end{eqnarray}
(Interchange of integration and limits is permitted by the monotone convergence theorem.)
 We denote the integral in (\ref{IN}) by
$I(N,\alpha)$;
it can be computed via integration on the cylinder set  defined earlier ({\it cf.}~(\ref{Rmeasure})) with ${\cal R} = {\bf R}^{2N}$. We have
\begin{equation}
\eqalign{
I(N,\alpha) &= \int_{{\bf R}^{2N}} \prod_{j=1}^NdQ_j dQ_{j}^* \cr
&\quad\quad\times{4\over\pi^2|H_0(|j|T)|^2} \exp\left\{-{4\over \pi}( \alpha + {1\over |H_0(|j|T)|^2})|Q_j|^2\right\}\cr
&=\prod_{j=1}^N {1\over 1+ \alpha  |H_0(|j|T)|^2}.
}
\end{equation}
Using the asymptotic behavior of the Hankel function it follows that 
\begin{equation}
\lim_{N\to\infty} I(N,\alpha) = 0
\end{equation}
so that $\mu_{\scriptscriptstyle T}(L^2({\bf S}^1))=0$. Thus we can say that ``typical'' configurations of the quantized Gowdy field $\hat\phi(T)$ are slightly more singular than square-integrable functions but are not as singular as, say, the delta function.

The preceding results are valid for any element of the family of measures $\mu_{\scriptscriptstyle T}$, $T>0$.  But this does not imply the support of the measure is the same for all time. It is not. More precisely, recall that two measures $\mu$ and $\mu^\prime$ on a space $M$ are said to be mutually singular if there exists a set $A\subset M$ such that $\mu(A)=0$ and $\mu^\prime(M\setminus A) = 0$. We shall now show that the Gaussian measures defined at two different times  are mutually singular. 

We will borrow methods from \cite{MTV}, which are designed to deduce finer support properties of Gaussian measures than can be obtained via the Minlos theorem. To begin with, it is straightforward to modify the proof of their Proposition 1 --- which deals with an infinite product of identical Gaussian measures --- to obtain the following result for the measure defined by (\ref{Cdefn}).  Let $\{\Delta_k\}$, $\Delta_k>1$, $k=1, 2, \dots$  be a given sequence.
Consider the set 
\begin{eqnarray}
Z_T(\Delta) = \{Q\in {\cal S}^\prime\, |\  &\exists\, N\in{\bf Z}^+\ {\rm such\ that}\ {\rm when\ }n>N\\ 
&|Q_n|<|H_0(|n|T)|\sqrt{{\pi\over4}\ln\Delta_n}\}.
\label{Z}
\end{eqnarray}
The $\mu_{\scriptscriptstyle T}$ measure of $Z_T(\Delta)$ is unity if
\begin{equation}
\sum_{k=1}^\infty {1\over \Delta_k\sqrt{\ln \Delta_k}} <\infty,
\label{Deltacondition}
\end{equation}
while the measure of $Z_T(\Delta)$ vanishes if the series diverges.

Next (still following the strategy of \cite{MTV}), we note the sequence $\Delta_n=n^\alpha$ satisfies (\ref{Deltacondition}) for any $\alpha>1$, while it fails to satisfy (\ref{Deltacondition}) for $\alpha=1$. Thus $\mu_{\scriptscriptstyle T}(Z_T(\{n^\alpha\})) = 1$ for $\alpha>1$ and the measure vanishes on the set with $\alpha=1$. 

Consider two times $T_1$ and $T_2$  with corresponding measures $\mu_{\scriptscriptstyle T_1}$ and $\mu_{\scriptscriptstyle T_2}$ . It is straightforward to verify that
\begin{equation}
Z_{T_1}(\{n\}) = Z_{T_2}(\{n^\xi\}),
\end{equation}
where
\begin{equation}
\xi = \left|{H_0(|n|T_1)\over H_0(|n|T_2)}\right|^2.
\end{equation}
As $n\to\infty$ we have
$$
\xi \sim {T_2\over T_1},
$$
so that $\Delta_n = n^\xi$ satisfies (\ref{Deltacondition}) for any $T_2>T_1$. We see then that 
\begin{equation}
\mu_{\scriptscriptstyle T_1}(Z_{T_1}(\{n\})) = 0, \quad \mu_{\scriptscriptstyle T_2}(Z_{T_1}(\{n\}))=1.
\end{equation}
It is also easy to see that
\begin{equation}
\mu_{\scriptscriptstyle T_2}(Z_{T_1}(\{n^\xi\}))=1
\end{equation}
It is now easy construct a set disjoint from $Z_{T_1}(\{n\})$ within $Z_{T_1}(\{n^\xi\}))$ which has  vanishing $\mu_{\scriptscriptstyle T_2}$ measure. Indeed, we have 
\begin{equation}
\mu_{\scriptscriptstyle T_2}\left(Z_{T_1}(\{n^\xi\})\setminus Z_{T_1}(\{n\})\right) = 0.
\end{equation}
\bigskip

\ack

This work was supported in part by National Science Foundation grant PHY-0244765 to Utah State University.

\end{document}